\documentclass[aps,prx,longbibliography,twocolumn,nofootinbib]{revtex4-2}

\usepackage[utf8]{inputenc}
\usepackage[T1]{fontenc}
\usepackage{amsmath}
\usepackage{amssymb}
\usepackage{amsthm}
\usepackage{mathtools}
\usepackage{enumitem}
\usepackage{graphicx}
\usepackage{hyperref}
\usepackage{seqsplit}
\usepackage{color}
\usepackage{cleveref}
\usepackage{braket}

\graphicspath{{figures/}}

\hypersetup{
  colorlinks=true,
  linkcolor=blue,
  citecolor=blue,
  urlcolor=blue
}

\newcommand{\tr}[2][]{\operatorname{Tr}_{#1}\!\left[#2\right]}

\newcommand{\IonQ}{\affiliation{%
IonQ Inc, College Park, MD, 20740, USA}}

\date{\today}
\begin{document}
\title{Accelerating Quantum-Classical Auxiliary-Field Quantum Monte Carlo with Shadow-Trained Neural Quantum States}
\author{Justin Provazza}
\email[Corresponding author: ]{justin.provazza@ionq.co}
\IonQ
\author{Oliver Knitter}
\thanks{J.P. and O.K. contributed equally to this work.}
\IonQ
\author{Joshua Goings}
\IonQ
\author{Evgeny Epifanovsky}
\IonQ
\author{Chi Chen}
\IonQ
\author{Martin Roetteler}
\IonQ
\author{Luning Zhao}
\IonQ

\begin{abstract}
Quantum-classical auxiliary-field quantum Monte Carlo (QC-AFQMC) uses a trial state prepared on a quantum computer to improve the accuracy of the classical AFQMC algorithm. State-of-the-art QC-AFQMC represents this state with matchgate shadows, so every trial-walker overlap and local energy must be averaged over many snapshots, making this classical computation a bottleneck. Here we distill the matchgate shadow data into a symmetry-constrained, noise-aware, autoregressive transformer neural quantum state (NQS). Used as the AFQMC trial state, the NQS evaluates overlaps, force biases, and local energies without repeatedly summing over the matchgate shadows. 
For a nickel-catalyzed oxidative-addition reaction, the noise-awareness of our NQS design successfully recovers the ideal-upCCD AFQMC barrier and reaction energy within statistical uncertainty, while standard QC-AFQMC execution with the same shadow data yields qualitatively inaccurate results as a result of hardware noise. Since the shadow data are needed only for training, the estimated classical cost for the nickel system falls by approximately 99\% relative to our previous QC-AFQMC implementation.
\end{abstract}

\maketitle
\section{Introduction}\label{sec:Introduction}
Modeling the electronic structure of chemical and materials systems is one of the most promising applications of a universal quantum computer. Accurate predictions of electronic properties support many use cases including drug discovery~\cite{Holzmann22_7001}, catalyst design~\cite{reiher17_7555, Troyer21_033055}, battery developments~\cite{Garcia21_134115, Zhao22_60}, and photovoltaic materials~\cite{Claudino23_5511}. The quantum-classical auxiliary-field quantum Monte Carlo (QC-AFQMC) algorithm provides a viable way to model electronic structure on near-term and early fault tolerant quantum computers. QC-AFQMC is a quantum extension of classical AFQMC~\cite{motta18_1364, lee22_7024, Zhao20_174105}, which propagates a stochastic representation of the wave function in imaginary time, converging in the long-time limit to the ground state~\cite{zhang97_7464,zhang03_136401,motta18_1364,lee22_7024,shee23_140901}.
In practice, a trial state is used to constrain the propagation and control a fundamental bottleneck for generic projection-based fermionic Monte Carlo methods, the sign problem~\cite{Umrigar143_164105, troyer05_170201}. However, this constraint introduces an energy bias that vanishes as the trial state approaches the exact eigenstate. A quantum computer that is able to efficiently prepare highly accurate, or nearly exact, trial states could benefit AFQMC by reducing or even eliminating this bias. 

The first demonstration of QC-AFQMC was performed in 2022~\cite{huggins22_416} when a trial state prepared on Google's Sycamore quantum computer was subsequently used to guide imaginary time propagation in AFQMC. Early results on molecular and lattice systems have further shown that quantum trial states can significantly reduce the energy bias relative to conventional classical trial states~\cite{huggins22_416, huang24_043063}, making QC-AFQMC a promising algorithm for quantum simulation. To use a quantum trial state in QC-AFQMC, however, one must repeatedly compute the overlap between the trial state $\ket{\Psi_T}$ and an arbitrary, non-orthogonal Slater determinant $\ket{\Phi}$, and state-of-the-art methods accomplish this task through shadow tomography~\cite{huang20_1050, Zhao21_110504, wan23_629}. Shadow tomography constructs a reusable classical representation from randomized one-shot measurements of the quantum trial state, allowing for simultaneous measurement estimates of $M$ target observables at $\mathcal{O}(\log M)$ cost~\cite{huang20_1050,Preskill22_2022}. The shadows are collected once and reused throughout the AFQMC workflow to estimate the necessary overlaps and local quantities, bypassing the need to repeatedly query the quantum device throughout imaginary time propagation.

Nonetheless, the computational cost for implementing the QC-AFQMC algorithm strongly depends on how efficiently one can post-process the shadow measurements to produce an overlap estimator~\cite{Kiser24_033022}, and significant advancements have been made toward improving this cost scaling since the invention of QC-AFQMC. Though initial work with Clifford shadows~\cite{huang20_1050} demonstrated an exponential scaling for overlap computation, the utilization of matchgate shadows~\cite{wan23_629} successfully yielded a polynomial scaling cost of $\widetilde{\mathcal{O}}(N^{4.5})$, where $N$ is the number of spin orbitals. A subsequent algorithm~\cite{nappi26_19239} further reduced this cost to $\widetilde{\mathcal{O}}(N^{3.5})$. Separately, the recently-developed orbital-rotation shadow protocol~\cite{low24_08964, zhao03103_2026} achieves a cost scaling of $\mathcal{O}(N\eta^3)$, where $\eta$ is the total number of electrons. These costs are for estimating a single overlap at fixed additive accuracy, and $\widetilde{\mathcal{O}}$ suppresses polylogarithmic factors. 

The aforementioned reduction in asymptotic scaling notwithstanding, QC-AFQMC faces a considerable efficiency issue: the shadow tomography imposes a substantial measurement overhead.  
Recent work applying QC-AFQMC to the oxidative addition step of a nickel-catalyzed Suzuki--Miyaura reaction using trial states prepared in an (8e, 8o) active space~\cite{zhao26_033061} produced $\sim$60,000 measurements for each molecular structure, and the computations for overlaps, force-biases, and local energies in QC-AFQMC explicitly iterated over all of them. The resulting cost not only grows with the number of shadows collected but is also incurred for every walker at every step throughout imaginary time propagation. Since the number of shadows required increases with system size, such an overhead severely limits the ability of QC-AFQMC to efficiently tackle large scale systems. However, the shadow measurements need not remain the final representation of the quantum trial state. Instead, they may be treated as a reusable source of complex overlap data from which a more compact wave function representation is learned~\cite{torlai18_447, wei24_023250,lange25_1675}.

Neural quantum states (NQSs)~\cite{carleo17_602} provide a natural representation for this purpose~\cite{choo20_2368}, since their model complexity is not explicitly tied to a physically motivated truncation of the wave function, and they can be fit directly to measurement data~\cite{lange24_040501}. NQSs are largely model-agnostic, with an indirect relationship between their
parameters and the physical structure of the states they represent~\cite{valenti22_L012010}. This offers great flexibility but can also make it difficult to systematically assess their performance. Design and training choices are typically determined by trial and error as one cannot identify \textit{a priori} the best architecture for a given problem, and optimization involves navigating a non-convex landscape without convergence guarantees~\cite{lange24_040501}. Techniques such as variational neural annealing~\cite{hibat21_952} (VNA) and the enforcement of symmetries~\cite{malyshev23_04166} can mitigate these optimization difficulties~\cite{lange24_040501} and, as discussed in Sec.~\ref{sec:nqs-train}, we employ both in our work.

An autoregressive NQS~\cite{sharir20_020503,barrett22_351}, by construction, encodes normalized complex amplitudes and enables efficient exact sampling from the corresponding probability distribution~\cite{zhao21_1}. Recently, NQS trial states have been incorporated directly into AFQMC through stochastic evaluation of the required mixed estimators~\cite{xiao25_9587}, with the dominant computational cost arising from energy minimization of the NQS serving as the trial state. This cost suggests an alternative QC-AFQMC workflow in which matchgate shadows are contracted with a set of representative determinants only once, and the resulting overlaps are distilled into an NQS for subsequent use throughout the production calculation.

\begin{figure}[t]
    \centering
    \includegraphics[width=\columnwidth]{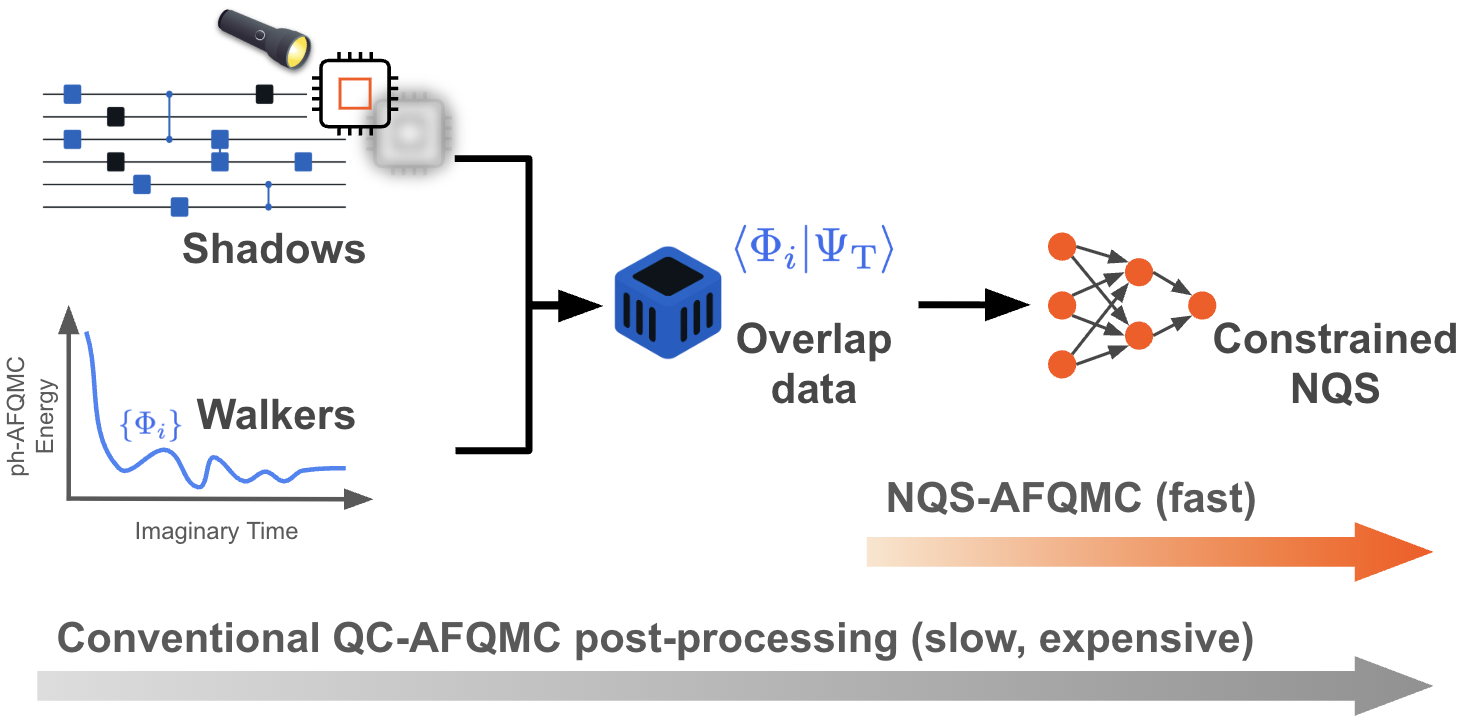}
    \caption{Workflow for constructing and using a shadow-trained NQS trial. A correlated trial state is prepared on the QPU and measured using matchgate shadows. Walkers $\{\Phi_i\}$ sampled from a preliminary ph-AFQMC calculation with a single Slater RHF trial are contracted with the shadow data to estimate the overlaps $\langle\Phi_i|\Psi_T\rangle$. The resulting supervised dataset is used to train a symmetry-constrained NQS with the calibrated and regularized objective in Eq.~\eqref{eq:calibrated-shadow-loss}. Conventional QC-AFQMC repeatedly processes the complete shadow data during propagation, whereas NQS-AFQMC transfers this dependence to the one-time construction of the training data and evaluates the learned trial during production.}
    \label{fig:shadow-nqs-pipeline}
\end{figure}

In this work, we develop and test a shadow-trained NQS representation for QC-AFQMC demonstrated to achieve a substantial post-processing cost reduction compared to conventional QC-AFQMC~\cite{zhao26_033061}. Here, shadow measurements are first used to compute overlaps with a collection of determinants sampled from a preliminary classical AFQMC (ph-AFQMC) run with a single Slater determinant trial state. These overlap--determinant pairs comprise a supervised dataset.  We then train committees of symmetry-constrained autoregressive NQSs and select a trial state from the committee in an objective and scalable manner. The selected state is subsequently integrated with AFQMC  through stochastic NQS sampling~\cite{xiao25_9587}. This approach moves the dependence on the number of shadow measurements to a one-time construction of the training data, removing it from production AFQMC post-processing. The construction of the overlap dataset and constrained NQS is summarized in Fig.~\ref{fig:shadow-nqs-pipeline}.

We first apply this approach to the nitrogen dissociation potential energy surface using a unitary pair coupled cluster doubles (upCCD)~\cite{Zhao22_60,Zhao24_76} trial state in a small active space. This experiment demonstrates how symmetry-constrained NQS training can approximate the effective trial wave function with high fidelity in the presence of hardware noise and finite shadow reconstruction error. We then apply this workflow using the same nickel oxidative-addition structures and QPU measurements considered in Ref.~\cite{zhao26_033061}, giving an estimated cost reduction of approximately 99\% in classical postprocessing while reproducing the ideal-upCCD AFQMC energetics within statistical uncertainty. We note that all results in this work utilize classically simulable active space sizes, enabling thorough analysis while also remaining within current quantum hardware constraints. Future work can extend this analysis to larger active spaces and more expressive ans\"atze on subsequent generations of trapped ion quantum computers.

The remainder of this paper is organized as follows. Section~\ref{sec:Theory} describes all relevant theory: the construction of walker-overlap training data from matchgate shadows, the calibrated NQS objective, and the deterministic and stochastic routes for evaluating the learned trial in AFQMC. Section~\ref{sec:comp-details} explains the computational aspects of our experimental setup. In Section~\ref{sec:Results}, we present the nitrogen and nickel oxidative-addition results alongside our analysis. We conclude in Section~\ref{sec:Conclusions} with a discussion of the scope and avenues to address the remaining limitations of shadow-trained trial states.

\section{Theory}\label{sec:Theory}

In this section, we describe the general theory underlying this architecture, the construction of training data from matchgate shadows, the NQS training pipeline, and the application of the trained NQS in AFQMC. We consider an electronic Hamiltonian expressed in an orthonormal spin-orbital basis,
\begin{equation}
    \hat{H}
    = \sum_{pq} h_{pq} a_p^\dagger a_q
    + \frac{1}{2}\sum_{pqrs} v_{pqrs}
      a_p^\dagger a_q^\dagger a_s a_r .
    \label{eq:electronic-hamiltonian}
\end{equation}
Auxiliary-field quantum Monte Carlo (AFQMC) projects the ground state by representing the imaginary time propagator as an ensemble of stochastic one-body propagators acting on \textit{walkers}~\cite{zhang03_136401}. Each walker is a Slater determinant $\ket{\Phi}$, and a trial state $\ket{\Psi_T}$ is used to control importance sampling and the phaseless constraint.  In particular, propagation requires the importance overlap
\begin{equation}
    \mathcal{A}_T(\Phi) = \langle \Psi_T | \Phi \rangle,
    \label{eq:trial-overlap}
\end{equation}
as well as ratios of this quantity and mixed estimates such as the local energy given by
\begin{equation}
    E_L(\Phi)
    = \frac{\langle \Psi_T | \hat{H} | \Phi \rangle}
           {\langle \Psi_T | \Phi \rangle}.
    \label{eq:local-energy}
\end{equation}
The accuracy of the phaseless approximation improves with the quality of the trial state, so the ability to prepare highly accurate trial states, potentially beyond what is classically tractable, would enable energy predictions of great accuracy. However, a classically intractable trial state is only useful in the context of AFQMC if Eqs.~\eqref{eq:trial-overlap} and \eqref{eq:local-energy} can be evaluated repeatedly for a large walker population, to support statistical convergence.

For $M$ spatial orbitals with $N_\alpha$ spin up and $N_\beta$ spin down electrons, we express a walker as
\begin{equation}
    \ket{\Phi}
    = \prod_{i=1}^{N_\alpha}
      \left(\sum_{p=1}^{M}\Phi^\alpha_{pi}a^\dagger_{p\alpha}\right)
      \prod_{j=1}^{N_\beta}
      \left(\sum_{q=1}^{M}\Phi^\beta_{qj}a^\dagger_{q\beta}\right)
      \ket{0}.
    \label{eq:walker-determinant}
\end{equation}
If $\ket{x}$ is an occupation number determinant and $X_x^\sigma$ is the
matrix whose columns select its occupied orbitals of spin $\sigma$, then
\begin{equation}
    \langle \Phi | x \rangle
    = \det\!\left[(\Phi^\alpha)^\dagger X_x^\alpha\right]
      \det\!\left[(\Phi^\beta)^\dagger X_x^\beta\right].
    \label{eq:determinant-overlap}
\end{equation}
This determinant structure permits polynomial-time evaluation of every computational-basis matrix element needed below.

\subsection{Training data from matchgate shadows}

Let $\ket{\Psi_T}$ denote an $\eta$-fermion trial state prepared on a quantum device.  Directly evaluating its overlap with every AFQMC walker would require new quantum measurements throughout the Monte Carlo calculation.  Instead, we measure the state once using classical matchgate shadows, subsequently reusing the shadow dataset to estimate overlaps with an arbitrary collection of Slater determinants~\cite{huggins22_416,wan23_629}. For the calculations considered here, the quantum state is prepared in a reduced active space $A$, and we write it as $\ket{\Psi_{T,A}}$ when distinguishing it from the corresponding full-space trial.

The complex phase of an overlap is made observable by preparing the coherent reference state,
\begin{equation}
    \ket{\tau} = \frac{1}{\sqrt{2}}
    \left(\ket{\mathrm{vac}} + \ket{\Psi_T}\right),
    \label{eq:bell-reference-state}
\end{equation}
where $\ket{\mathrm{vac}}=\ket{0\cdots0}$ is the vacuum state. For each snapshot $n$, a randomly selected fermionic Gaussian unitary $U_n$ is applied before measurement in the occupation basis, producing a bit string $b_n$.  If $\mathcal{M}$ denotes the corresponding average measurement channel, the single-shot classical snapshot is
\begin{equation}
    \hat{\rho}_n
    = \mathcal{M}^{-1}\!\left[
        U_n^\dagger\ket{b_n}\!\bra{b_n}U_n
      \right].
    \label{eq:classical-snapshot}
\end{equation}
The overlap with a query determinant is estimated from the off-diagonal element
of $\rho_\tau=\ket{\tau}\!\bra{\tau}$,
\begin{equation}
    \hat{y}_i
    = \frac{2}{N_{\mathrm{sh}}}\sum_{n=1}^{N_{\mathrm{sh}}}
      \tr{\ket{\mathrm{vac}}\!\bra{\Phi_i}\hat{\rho}_n}
    \approx \langle \Phi_i | \Psi_T \rangle.
    \label{eq:shadow-overlap-estimator}
\end{equation}
The vacuum reference fixes the global phase of the complex labels. Because each matchgate acts as an orthogonal transformation on the Majorana operators, the traces in Eq.~\eqref{eq:shadow-overlap-estimator} reduce to covariance-matrix operations and Pfaffians~\cite{wan23_629}.

The determinants used to construct the training set are sampled from preliminary ph-AFQMC walker trajectories in the full molecular-orbital space. For the applications described here, these calculations are initialized from the RHF determinant that also serves as a trial state. Before evaluating a shadow overlap, each full-space walker $\Phi_i$ is first projected to the active space. This is done with the virtual correlation energy (VCE) technique.  With VCE, a full space determinant is reduced to a product of an orthonormal active-space determinant $\widetilde{\Phi}_{i,A}$ and a complex factor $\mathcal{N}(\Phi_i)$,
\begin{equation}
    \langle \Phi_i|\Psi_T\rangle
    = \mathcal{N}(\Phi_i)
      \langle\widetilde{\Phi}_{i,A}|\Psi_{T,A}\rangle.
    \label{eq:active-space-overlap}
\end{equation}
The factor $\mathcal{N}(\Phi_i)$ contains the normalization and phase acquired when the frozen occupied and active blocks of the walker are separated. We then discard it and only use the active space overlap $\hat{y}_i\approx\langle\widetilde{\Phi}_{i,A}|\Psi_{T,A}\rangle$ to construct the dataset. Equation~\eqref{eq:shadow-overlap-estimator} is therefore evaluated using $\widetilde{\Phi}_{i,A}$ as the query determinant.  The same shadow data are reused for every query, and the final supervised data are the active-space pairs $\{\widetilde{\Phi}_{i,A},\hat{y}_i\}_{i=1}^{N_D}$.  These pairs remain focused on the determinant manifold visited by the downstream AFQMC calculation without requiring the NQS to represent the frozen or external orbitals. To simplify the notation below, we denote the stored active-space determinants as $\Phi_i$ after this point.

The NQS described in the next subsection consequently learns only the active-space state. Meanwhile, AFQMC works in the full orbital space. To address this mismatch, each exported configuration is embedded back into the full orbital space by padding it with doubly occupied frozen orbitals and empty external virtual orbitals,
\begin{equation}
    \ket{\Psi_T}
    = \ket{\Xi_c^\alpha}\ket{\Xi_c^\beta}
      \ket{\Psi_{\theta,A}}
      \ket{0_v^\alpha}\ket{0_v^\beta}.
    \label{eq:active-space-trial}
\end{equation}
The learned part of the wave function therefore remains confined to the quantum active space.  The padded configurations are evaluated against the full-space walkers using the stochastic NQS trial described below, while propagation with the full-space Hamiltonian recovers external correlation.

\subsection{Shadow-trained neural quantum state}
\label{sec:nqs-train}

\begin{figure}[t]
    \centering
    \includegraphics[width=0.95\columnwidth]{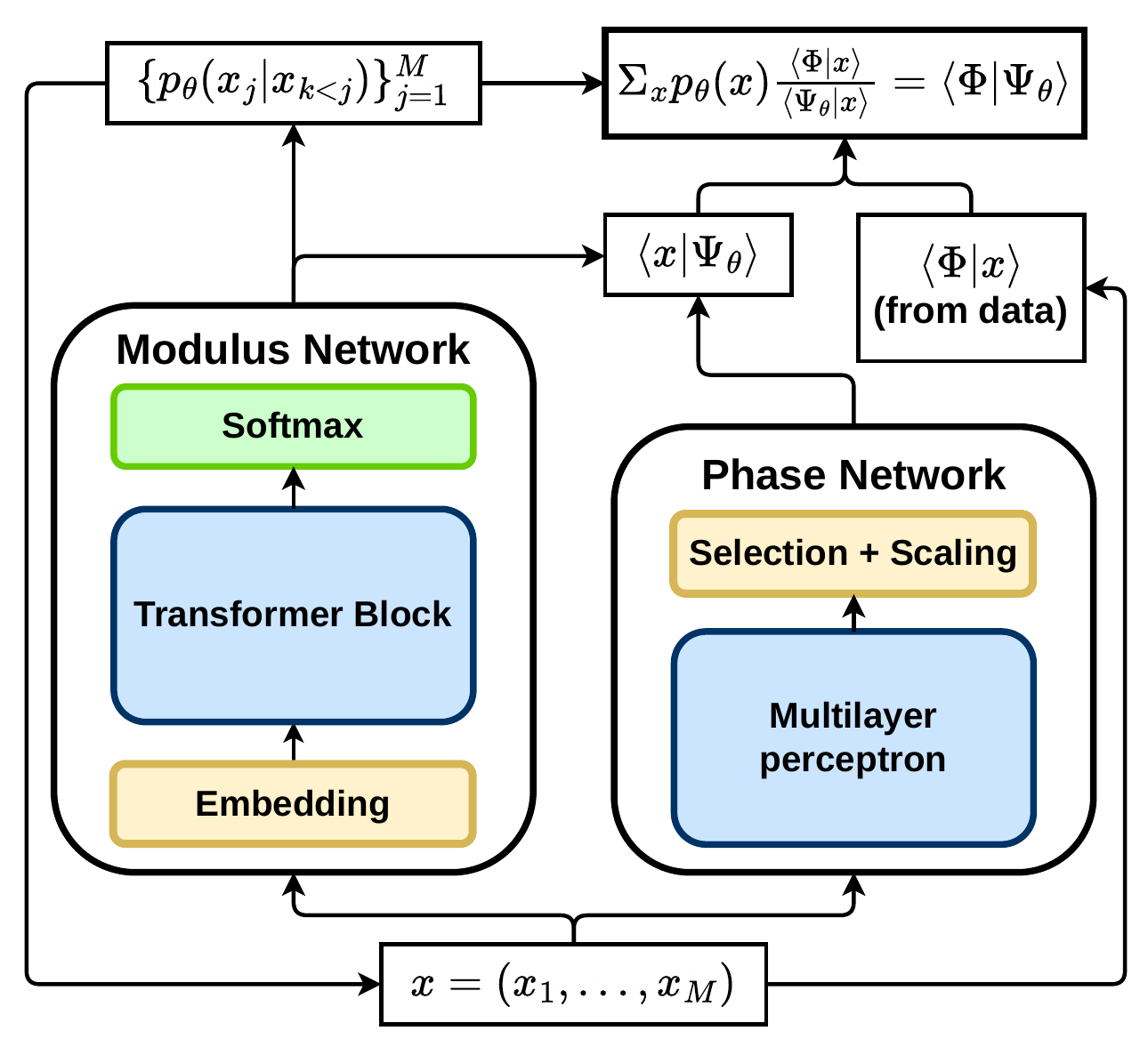}
    \caption{The autoregressive NQS architecture considered in this work comprises a modulus and phase network operating in tandem to encode statevector entries of $\ket{\Psi_\theta}$. The outputs of these networks are used to produce stochastic estimators of overlaps with the determinants from the training data.
    }
    \label{fig:nqs_diagram}
\end{figure}

We represent the trial state by using an autoregressive NQS~\cite{carleo17_602,sharir20_020503},
\begin{equation}
    \ket{\Psi_\theta}
    = \sum_x \psi_\theta(x)\ket{x},
    \qquad
    \psi_\theta(x)
    = \sqrt{p_\theta(x)}e^{i\varphi_\theta(x)}.
    \label{eq:nqs-state}
\end{equation}
The occupation of each spatial orbital is treated as a four-valued variable, $x_j\in\{0,\alpha,\beta,\alpha\beta\}$, and the probability factorizes as
\begin{equation}
    p_\theta(x) = \prod_{j=1}^{M}
    p_\theta(x_j|x_1,\ldots,x_{j-1}).
    \label{eq:autoregressive-factorization}
\end{equation}
This construction defines a normalized state and permits exact independent sampling from $p_\theta(x)=|\psi_\theta(x)|^2$ through orbital-by-orbital sampling of the nested conditional probabilities. These probabilities are encoded in an autoregressive modulus network~\cite{sharir20_020503,barrett22_351}, while a phase network $\varphi_\theta$ maps global orbital occupancy configurations to phase values. We follow the paradigm that uses separate neural networks to model the modulus and phase~\cite{zhao23_025034}, but alternate architectures do exist that use modulus network output to construct phases~\cite{bennewitz22_618}. The two-network setup typically utilizes a shallow multilayer perceptron for $\varphi_\theta$, while the autoregressive modulus network requires a more sophisticated design. We specifically focus on the transformer-based architecture~\cite{bennewitz22_618,wu23_1,shang25_8464}, a central subject of analysis within the wider NQS literature~\cite{knitter25_025022,xiao25_9587}. Figure~\ref{fig:nqs_diagram} gives an overview of this architecture and how it is used to produce overlaps with potentially arbitrary determinants. The architecture and optimization parameters used in the present calculations are provided in Sec.~\ref{app:nqs-training}.

Although the network is evaluated on occupation configurations, the supervised training targets are overlaps with generally nonorthogonal determinants.  Combining Eqs.~\eqref{eq:determinant-overlap} and \eqref{eq:nqs-state} gives
\begin{align}
    \langle \Phi | \Psi_\theta \rangle
    &= \sum_x \langle\Phi|x\rangle\psi_\theta(x) \nonumber\\
    &= \mathbb{E}_{x\sim p_\theta}
      \left[\frac{\langle\Phi|x\rangle}{\psi_\theta^*(x)}\right].
    \label{eq:nqs-overlap-expectation}
\end{align}
Given a set of $N_s$ autoregressive samples $x_k\sim p_\theta$, we use the stochastic
estimator
\begin{equation}
    \langle\Phi|\Psi_\theta\rangle
    \simeq \frac{1}{N_s}\sum_{k=1}^{N_s}
      \frac{\langle\Phi|x_k\rangle}{\psi_\theta^*(x_k)}.
    \label{eq:nqs-overlap-estimator}
\end{equation}
The same NQS samples can be shared across a batch of training determinants, so the determinant-overlap evaluations are naturally vectorized.

Physical symmetry constraints can be imposed directly on the autoregressive NQS by masking the modulus network's conditional probability outputs during the sampling process~\cite{malyshev23_04166}, such that forbidden configurations have $p_\theta(x)=0$. For example, fixing $(N_\alpha,N_\beta)$ confines the NQS to a definite particle number and spin projection $S_z=(N_\alpha-N_\beta)/2$. Other masks corresponding with desirable wave function structures may be applied similarly. Consequently, no separate constraint penalty is required in the training loss as the restriction enters through the support of $\ket{\Psi_\theta}$, ensuring every sample used to estimate a walker overlap lies within the selected sector. 

A dataset constructed with noiseless shadows preserves the symmetry of the trial state, and a constrained NQS trained on it should also preserve that symmetry. However, when using noisy shadows obtained on quantum devices, the dataset itself becomes noisy and may break the symmetry. A symmetry constrained NQS that is not aware of noise will then randomly impose the constraint on the noisy data, so naively training an NQS in this way can lead to a significant mismatch between the trained state and the actual trial state. 

To deal with this, we introduce a positive noise calibration factor $a$ in the loss function.   
We make the simplest assumption that noise predominantly manifests as a common multiplicative scaling, absorbing any known dataset-wide rescaling, including parity conditioning when it acts as a common rescaling of all labels. A Bayesian derivation and its determinant-dependent generalization are discussed in Appendix~\ref{app:calibration}, along with a description of noise models under which this uniform response is exact in expectation. Treating $a$ as learnable and adding a regularization term to prevent overfitting, we minimize the loss function
\begin{equation}
    \mathcal{L}(\theta,a)
    = \frac{1}{N_D}\sum_{i=1}^{N_D}
      \left|a\,\langle\Phi_i|\Psi_\theta\rangle-\hat{y}_i\right|^2+\lambda a^2.
    \label{eq:calibrated-shadow-loss}
\end{equation}

The NQS overlaps in Eq.~\eqref{eq:calibrated-shadow-loss} are evaluated stochastically via Eq.~\eqref{eq:nqs-overlap-estimator}. In contrast with the neural shadow tomography workflow of Ref.~\cite{wei24_023250}, which minimizes an infidelity estimated from classical shadows, this objective directly regresses the NQS-predicted walker overlaps against overlaps estimated from the shadows. Note that the uncalibrated objective is recovered simply by setting $a=1$ and $\lambda=0$. The easiest way to understand the motivation for introducing such a multiplicative factor is to consider it as the noise calibration coefficient. With noisy shadows, the computed overlap amplitudes can be suppressed, for example, by population leakage into the symmetry violating subspace, e.g. states with wrong $S_z$ symmetry or incorrect particle numbers. The factor $a$ then attempts to accommodate for this by scaling NQS overlaps.

The ridge term in Eq.~\eqref{eq:calibrated-shadow-loss} acts only on the single calibration scalar and is distinct from the AdamW weight decay regularizing the neural network parameters $\theta$. We disable this optimizer weight decay for the unconstrained parameter representing $a$. We include VNA~\cite{hibat21_952} during early training, which has been shown to improve NQS performance when directly optimizing chemistry Hamiltonian energies~\cite{knitter25_025022}. VNA regularizes the loss function by $\kappa_t\sum_x p_\theta(x)\log p_\theta(x)$ and gently encourages the NQS toward higher-entropy states in order to prevent early model collapse. Similar to what was done in Ref.~\cite{knitter25_025022}, the regularization coefficient $\kappa_t$ is reduced to 0 along a quintic schedule beginning at $5\%$ of the total number of optimization steps.
Thus $\lambda$, AdamW weight decay, and the annealing schedule control three different aspects of the optimization.

At each optimization step, the same autoregressive samples are shared across each batch of walker determinants. Following the importance sampling gradient optimization (ISGO) method~\cite{yang20_012039}, we further reuse samples between steps by reweighting the terms in Eq.~\ref{eq:nqs-overlap-estimator} using importance factors that correct for the difference in sampling distributions between optimization steps, refreshing the sample cache when these factors exceed a prescribed threshold.  For the cross-run comparison reported below, every optimization is evaluated from the latest saved checkpoint at the same training epoch.  Exact-state fidelities and energies are reserved solely for auditing, never training. For the system sizes considered, we may query the exported model checkpoint to obtain the learned complex statevector amplitudes corresponding with every configuration allowable under the desired model constraint, with zero coefficient cutoff. The resulting coefficients are normalized for downstream use, which is consistent with the global scale--invariant use of overlap ratios in AFQMC. Because training probes only a finite, walker-derived determinant distribution, the procedure reconstructs the trial where it is needed for AFQMC and does not by itself imply globally complete tomography.

\subsection{Evaluation of the NQS trial in AFQMC}
\label{sec:nqs-afqmc}
The NQS trial wave function is integrated with AFQMC with stochastic sampling of the trial wave function. The stochastic NQS-AFQMC formulation closely follows Ref.~\cite{xiao25_9587}. For each walker, we sample the configuration distribution
\begin{equation}
    q_\Phi(x)=\frac{|\psi_\theta^*(x)\langle x|\Phi\rangle|}{Z_\Phi},
    \qquad
    Z_\Phi=\sum_x|\psi_\theta^*(x)\langle x|\Phi\rangle|,
    \label{eq:walker-importance-distribution}
\end{equation}
using a Metropolis chain over configurations sorted by $|\psi_\theta(x)|$. The normalization $Z_\Phi$ is not evaluated directly because it cancels in every ratio required by AFQMC. For $P$ samples $x_k\sim q_\Phi$, we define the unit-modulus phase factor
\begin{equation}
    S_\Phi(x)
    = \frac{\psi_\theta^*(x)\langle x|\Phi\rangle}
           {|\psi_\theta^*(x)\langle x|\Phi\rangle|},
    \label{eq:phase-factor}
\end{equation}
and samples drawn for a pre-propagation walker $\Phi$ give
\begin{equation}
    \frac{\mathcal{A}_\theta(\Phi')}{\mathcal{A}_\theta(\Phi)}
    \approx
    \frac{\displaystyle\sum_{k=1}^{P}
      \frac{\langle x_k|\Phi'\rangle}{\langle x_k|\Phi\rangle}S_\Phi(x_k)}
      {\displaystyle\sum_{k=1}^{P}S_\Phi(x_k)},
    \label{eq:nqs-overlap-ratio}
\end{equation}
where $\mathcal{A}_\theta(\Phi)=\langle\Psi_\theta|\Phi\rangle$. For a general operator $\hat{O}$, the corresponding estimator is given by
\begin{equation}
    \frac{\langle\Psi_\theta|\hat{O}|\Phi'\rangle}
         {\langle\Psi_\theta|\Phi\rangle}
    \approx
    \frac{\displaystyle\sum_{k=1}^{P}
      \frac{\langle x_k|\hat{O}|\Phi'\rangle}{\langle x_k|\Phi\rangle}
      S_\Phi(x_k)}
      {\displaystyle\sum_{k=1}^{P}S_\Phi(x_k)}.
    \label{eq:nqs-mixed-estimator}
\end{equation}
Thus the Green's function and local energy are evaluated as
\begin{align}
    G^\sigma_{pq}(\Phi)
    &\approx
    \frac{\sum_k S_\Phi(x_k)G^{\sigma,(x_k)}_{pq}(\Phi)}
         {\sum_k S_\Phi(x_k)},
    \label{eq:nqs-greens-function}\\
    E_L(\Phi)
    &\approx
    \frac{\sum_k S_\Phi(x_k)E_L^{(x_k)}(\Phi)}
         {\sum_k S_\Phi(x_k)},
    \label{eq:nqs-local-energy}
\end{align}
where the superscript $(x_k)$ denotes the conventional single-determinant quantity.  The force bias is obtained from the same phase-weighted Green's function contracted with the Cholesky vectors of the two-electron interaction.  The implementation aggregates repeated configuration labels, applies Sherman--Morrison--Woodbury updates to proposed determinant overlaps, and batches the Green's function, force bias, and local energy contractions over the unique sampled configurations.

The samples and walkers are advanced using a three-stage leapfrog procedure. First, the sample set $\mathcal{X}_{\Phi}$ is held fixed while $\Phi$ is propagated to $\Phi'$ and the walker weight is updated using Eq.~\eqref{eq:nqs-overlap-ratio}. Second, Metropolis sweeps update $\mathcal{X}_{\Phi}\rightarrow\mathcal{X}_{\Phi'}$ against the new walker. Finally, changing the stochastic representation at fixed $\Phi'$ introduces a hand-off ratio
\begin{equation}
\begin{aligned}
    R_{\mathrm{ho}}
    &= \frac{\sum_{x\in\mathcal{X}_{\Phi'}}S_{\Phi'}(x)}
            {\sum_{x\in\mathcal{X}_{\Phi}}S_{\Phi'}(x)},\\
    W'&\longrightarrow W'\max\!\left[
        0,\cos\!\left(\arg R_{\mathrm{ho}}\right)\right].
\end{aligned}
    \label{eq:nqs-handoff}
\end{equation}
Taken together, the estimators preserve a single underlying NQS trial across propagation, force bias, and energy measurement.

\section{Computational Details}
\label{sec:comp-details}
\subsection{Neural network training protocol}
\label{app:nqs-training}

All calculations use the same NQS architecture, with the input dimension determined by the active space. Each spatial orbital is represented by one of four occupation tokens corresponding to empty, $\alpha$-, $\beta$-, and doubly-occupied configurations. The order in which the modulus network operates on these orbitals is reversed from how they are considered by the second quantization. The modulus network itself comprises a single causal transformer block with an embedding dimension of 16, two attention heads, and a feed-forward dimension of 64. Following~\cite{wu23_1} we apply both learned token and positional embeddings to the input sequence, and we use the conditional masking procedure discussed in Sec.~\ref{sec:nqs-train} to restrict the NQS to the support defined by $(N_\alpha, N_\beta)$. Likewise, restriction to upCCD support was achieved by seniority-zero masking of the conditional probabilities.

The phase comprises a multilayer perceptron with two 64-neuron hidden layers and ReLU activations. It maps a global spatial orbital occupancy configuration, without the final orbital in the autoregressive ordering, to four outputs corresponding to the possible occupations of the final orbital. The selected output is multiplied by $\pi$ to obtain the phase of the configuration. For six- and eight-orbital active spaces, the NQS respectively contains 12,808 and 13,096 parameters, with one additional parameter needed for the learnable calibration factor. Model and calibration parameters are optimized with AdamW using a constant learning rate of $2.5\times10^{-3}$, $\beta_1=0.9$, $\beta_2=0.999$, $\epsilon=10^{-8}$, and a weight decay of $10^{-2}$ applied to the model parameters only. We also clip the NQS gradient to unit norm. Transformer linear and embedding weights are initialized from a normal distribution with $\sigma=0.02$, while the remaining weights are initialized according to PyTorch defaults~\cite{paszke19_8026}.

We geometrically increase the number of non-unique stochastic samples generated at each step from $10^2$ to $10^8$ over the course of training. The sampling procedure only explicitly stores the unique samples and keeps track of their multiplicities, retaining at most 10,000 unique configurations per sample batch generation. ISGO sample cache refresh is triggered whenever the largest absolute change in log probability exceeds 0.5. Validation losses are calculated after every epoch using a fresh NQS sample set, and both the lowest-validation-error and final checkpoints are retained for subsequent analysis.

Independent NQS optimizations can converge to distinct states even when they give similar training errors, so selection from a single optimization run introduces an uncontrolled random seed dependence. We thus train an ensemble of three NQSs for each value of the ridge parameter $\lambda$ using independent seeds. For production state selection, we first retain only ensembles for which the fidelity between every pair of normalized NQSs satisfies \begin{equation}
    \min_{i,j\in\{1,2,3\}}\left\lvert\left\langle\Psi^i_\theta\Big|\Psi^j_\theta\right\rangle\right\rvert^2\geq0.99.
\end{equation} We then select the individual normalized NQS with the lowest Rayleigh energy from all members of the retained committees. The energy can be evaluated directly for a sparse state or estimated stochastically for a larger NQS, so this criterion remains scalable.

\subsection{AFQMC simulation protocol}

Production NQS-AFQMC calculations for both applications used 1,280 walkers initialized from the RHF determinant, a time step of $\Delta\tau=0.005~E_h^{-1}$, and 20 propagation steps per energy block. Each block therefore spans $0.1~E_h^{-1}$ of imaginary time. Population control and walker stabilization were each applied every five propagation steps. We used $P=800$ NQS samples per walker for each NQS-AFQMC simulation. The corresponding Markov chains were initially thermalized with 1,000 Metropolis burn-in steps and were then advanced by 25 MCMC sweeps after each AFQMC propagation step. The reported energies are arithmetic means of the retained block estimates. Because successive block energies are correlated, we estimated the autocorrelation time for each trajectory and used it to select an appropriate reblocking length~\cite{malone23_109}. The reported one-standard-error uncertainties were calculated from these reblocked data. Additional details of the stochastic NQS-AFQMC implementation are provided in Ref.~\cite{xiao25_9587}.

\subsection{Dataset Construction}

To generate the training dataset, for each molecular structure, we first prepare the upCCD state in the selected active space, and obtain matchgate shadow measurements. We then run a classical ph-AFQMC calculation with 1024 walkers. We take snapshots of the walkers along the imaginary time propagation, and project them into the active space with VCE. In this way we build a collection of active space walkers: $\{\Phi_i\}$, after which matchgate shadows are used to compute the walker overlap with the trial state $y_i=\braket{\Phi_i|\Psi_T}$ as per Eq.~\eqref{eq:shadow-overlap-estimator}, and the overall dataset consists of pairs of walkers and overlaps: $\{\Phi_i\rightarrow y_i\}$. The constructed dataset contains 10,240 data points for N$_2$ and 154,624 data points for the Ni complexes. 

\section{Results}\label{sec:Results}

We first consider three points on the N$_2$ potential energy surface as a smaller controlled example in which the effect of the calibration and support constraint can be isolated.  We then apply the shadow-trained NQS to the nickel-catalyzed oxidative addition considered in Ref.~\cite{zhao26_033061}, which permits a direct comparison with the original matchgate-shadow workflow using the same quantum measurements and molecular Hamiltonians. 

\subsection{N$_2$ dissociation}
\label{sec:nitrogen}
The N$_2$ dissociation curve provides a controlled setting in which to examine the approach across distinct correlation regimes. All N$_2$ calculations use the 6-31G basis set. Around the equilibrium configuration, where the bond length is $r\approx1.1~\text{\AA}$, the molecule is described effectively \textit{via} single reference methods. However, as the bond is stretched toward the dissociative regime, a multireference description becomes necessary. To assess the abilities of the shadow-trained NQS scheme across different degrees of multireference character, we consider points at $r=1.1$, $1.6$, and $2.0~\text{\AA}$ along the N$_2$ stretch. 

For each of these 3 geometries, we prepare a upCCD state in the (6e,6o) active space around the Fermi level. We then measure matchgate shadows of the prepared state, on both an ideal simulator and the IonQ Forte Enterprise quantum computer. To understand the results, we first write a generic full configuration interaction wave function.
\begin{equation}
    \label{eqn:ideal_upccd_state}
    \ket{\Psi}=\sum_{i}^{N_{\mathrm{pair}}}{c_i\ket{\Phi^{\mathrm{pair}}_i}}+\sum_{j}^{N_{\mathrm{non-pair}}}{k_j\ket{\Phi^{\mathrm{non-pair}}_j}}
\end{equation}
where $\ket{\Phi^{\mathrm{pair}}_i}$ is a determinant in the seniority-zero space in which electrons form pairs and $\ket{\Phi^{\mathrm{non-pair}}_j}$ is a determinant outside the seniority-zero space, in which at least one electron pair is broken. 
For the (6e,6o) active space with no spin polarization, we find that $N_{\mathrm{pair}}=20$ and $N_{\mathrm{pair}}+N_{\mathrm{non-pair}}=400$. The system is small enough and it allows us to exactly reconstruct the state by computing the coefficients $c_i$ and $k_j$ using matchgate shadows as overlaps between the trial state and the corresponding determinant. We can then write the shadow-constructed state as

\begin{equation}
    \label{eqn:shadow_constructed_state}
    \ket{\Psi_s}=\frac{1}{\mathcal{N}}\left(\sum_{i}^{N_{\mathrm{pair}}}{c^s_i\ket{\Phi^{\mathrm{pair}}_i}}+\sum_{j}^{N_{\mathrm{non-pair}}}{k^s_j\ket{\Phi^{\mathrm{non-pair}}_j}}\right)
\end{equation}
in which $c^s_i$ and $k^s_j$ are computed using matchgate shadows.  We also add a normalization factor $\mathcal{N}$ since the shadow computed coefficients are not necessarily normalized. For the ideal upCCD state, all $k^s_j$ should be zero which is satisfied only if the number of (ideal) shadows goes to infinity. With a finite number of shadows or shadows collected on a noisy quantum hardware, $k^s_j$ could be nonzero and $c^s_i$ may deviate from the exact value. 

\begin{figure}[t]
    \centering
    \includegraphics[width=\columnwidth]{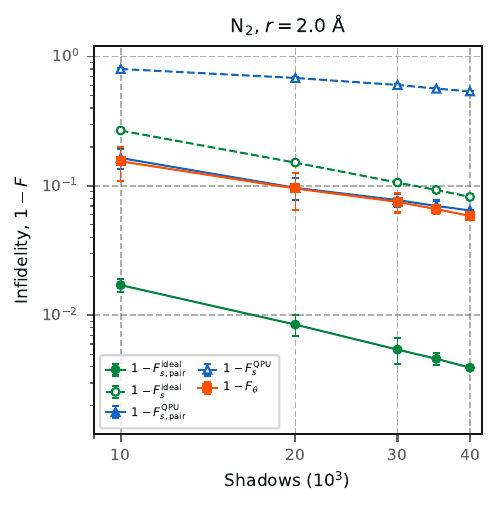}
    \caption{Matchgate shadow convergence and constrained NQS construction for N$_2$ at $r=2.0~\text{\AA}$. Infidelities are shown for the ideal simulator and QPU reconstructions on the seniority-zero space and in the full seniority space, together with the constrained NQS trained from the QPU shadow data. Each point below 40,000 shadows is a mean over ten matched subsets sampled without replacement, and the error bars represent one sample standard deviation.}
    \label{fig:n2-shadow-fidelity}
\end{figure}

We can also construct the following state
\begin{equation}
    \label{eqn:shadow_constructed_pair_state}
    \ket{\Psi_{s,\mathrm{pair}}}=\frac{1}{\mathcal{N}_{\mathrm{pair}}}\sum_{i}^{N_{\mathrm{pair}}}{c^s_i\ket{\Phi^{\mathrm{pair}}_i}}
\end{equation}
by removing all the non-pair determinants and re-normalizing. From the above three states and the trained NQS, $\ket{\Psi_{\theta}}$, we define the fidelity
\begin{equation}
    \begin{split}
        F^{X}_s&=|\braket{\Psi|\Psi_s}|^2 \\
        F^{X}_{s,\mathrm{pair}}&=|\braket{\Psi|\Psi_{s,\mathrm{pair}}}|^2 \\
        F_{\theta}&=|\braket{\Psi|\Psi_{\theta}}|^2,
    \end{split}
\end{equation}
and we denote fidelity computed using ideal (QPU) shadows as $F^{\mathrm{ideal}}_s$($F^{\mathrm{QPU}}_s$). For the current application, we take $\ket{\Psi}$ as the exact upCCD state. We plot the infidelity data ($1-F$) comparing ideal and QPU shadows  in Figure~\ref{fig:n2-shadow-fidelity} for $r=2.0~\text{\AA}$. 

The ideal simulator curves separate finite shadow reconstruction error from hardware noise. The difference between the two ideal simulator curves reflects leakage introduced by the finite number of shadows. Increasing the number of shadows improves both fidelities, and after removing the contribution from non-paired determinants, the state reconstructed from 40,000 ideal shadows is very close to the exact upCCD state. Hardware noise shifts both curves toward larger infidelity: For 40,000 shadows, $F^\text{QPU}_{s}=0.46$. After projection and renormalization on the paired only subspace, the reconstructed state nevertheless retains a fidelity of $F^\text{QPU}_{s,\text{pair}} = 0.94$. This indicates that the amplitude and phase structure within the paired determinant space is largely preserved.

Having characterized the effects of noise, we now seek to approximate the ideal upCCD trial state by training an NQS constrained to the seniority-zero subspace using QPU shadows. Although the NQS contains only paired configurations, the training walkers generally contain both paired and non-paired components, so their shadow-derived overlap labels also probe contributions outside the assumed NQS support. The training must therefore recover the paired component from these noisy overlap data. The NQS and $\ket{\Psi_{s,\mathrm{pair}}}$ are constructed from exactly the same shadows, and their infidelities are in excellent agreement within statistical uncertainty across the displayed shadow budgets. The constrained NQS reaches a fidelity of $F_\theta=0.94$ with 40,000 shadows.

\begin{table*}[t]
\caption{Absolute N$_2$ energies in $E_h$ using the 6-31G basis. The CASSCF+NEVPT2 results use the (6e,6o) active space.}
\label{tab:n2-afqmc}
\centering
\footnotesize
\setlength{\tabcolsep}{7pt}
\begin{ruledtabular}
\begin{tabular}{lccc}
Method & $1.1~\text{\AA}$ & $1.6~\text{\AA}$ & $2.0~\text{\AA}$ \\
\hline
Ideal-upCCD AFQMC & $-109.1071\pm0.0007$ & $-108.9504\pm0.0008$ & $-108.8238\pm0.0007$ \\
NQS-AFQMC & $-109.1048\pm0.0007$ & $-108.9423\pm0.0009$ & $-108.8064\pm0.0007$ \\
RHF & $-108.8676$ & $-108.5516$ & $-108.3096$ \\
CASSCF+NEVPT2 & $-109.0837$ & $-108.9211$ & $-108.8389$ \\
\end{tabular}
\end{ruledtabular}
\end{table*}

\begin{figure}[t]
    \centering
    \includegraphics[width=\columnwidth]{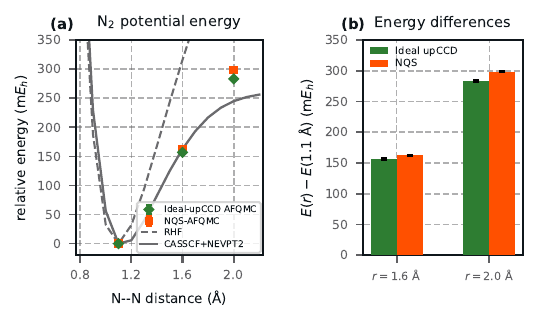}
    \caption{N$_2$ energetics in the 6-31G basis. (a) Relative energies, with each method referenced to its own value at $r=1.1~\text{\AA}$. Green diamonds and orange squares show Ideal-upCCD AFQMC and NQS-AFQMC, respectively. Dashed and solid gray curves show RHF and $(6e,6o)$ CASSCF+NEVPT2. Error bars on the AFQMC points are the one-standard-error uncertainties of the corresponding absolute energy estimates. (b) Energy differences $E(r)-E(1.1~\text{\AA})$ at $r=1.6$ and $2.0~\text{\AA}$ for the Ideal-upCCD and NQS trial wave functions, shown from left to right.}
    \label{fig:n2-energetics}
\end{figure}

We now use the trained NQS states as trials in AFQMC. For both the NQS and ideal-upCCD trials, each calculation was propagated for 600 blocks of 20 steps, corresponding to 12,000 propagation steps and a total imaginary time of $60~E_h^{-1}$. The first 300 blocks, corresponding to $30~E_h^{-1}$, were discarded to allow the walker population to equilibrate from the RHF initialization. The reported energies were obtained from the remaining 300 blocks and are shown in Table~\ref{tab:n2-afqmc}. Figure~\ref{fig:n2-energetics} shows the resulting relative-energy profiles, referenced to the energy of each method at $r=1.1~\text{\AA}$. The ideal upCCD AFQMC results are obtained by exactly computing the overlaps between the upCCD state and the AFQMC walkers (with exponential cost scaling). One could consider it as the noiseless, infinite-shadow limit. We also include the prediction of restricted Hartree-Fock (RHF) and the complete active space self-consistent field (CASSCF) + the N-electron valence state perturbation theory at the second order (NEVPT2) for reference. The CASSCF uses the same (6e,6o) active space as the upCCD state.

Looking at the absolute energy, we find that the NQS-AFQMC trained with QPU shadows yields energies slightly above the prediction of QC-AFQMC with the ideal upCCD trial state, with energy differences of 2.3, 8.1, and 17.4~m$E_h$ for $r=1.1$, $1.6$, and $2.0$~\AA, respectively. We do notice that the energy difference tends to increase with bond distance. This may reflect the imperfect NQS reconstruction of the upCCD overlaps due to shadows measured on noisy quantum devices, and the fact that strongly correlated geometries are especially sensitive to the quality of the trial states. However, both trial-state reconstruction and Monte Carlo evaluation of the NQS overlaps and mixed estimators during AFQMC propagation can contribute to the energy difference, and we do not disentangle their contributions here. 

One may also note that even with the ideal upCCD state, the relative energy of QC-AFQMC still differs from the prediction of CASSCF+NEVPT2 at stretched geometries. Specifically, the energy difference between $1.1~\text{\AA}$ and $2.0~\text{\AA}$ bond lengths predicted by QC-AFQMC is overestimated. This may reflect the quality of the trial state. At stretched geometries, the paired approximation made in upCCD may not be sufficient to capture the correlations and one may need to also include non-paired excitations in the trial state. 

\subsection{Nickel-catalyzed oxidative addition}
\label{sec:nickel}

Our next example is the oxidative addition step of the nickel-catalyzed deformylative Suzuki-Miyaura cross-coupling reaction~\cite{guo19_1957}. This is the same reaction used in our previous study~\cite{zhao26_033061}. The three structures considered here are the reactant B, transition state $[\mathrm{B-C}]^\ddagger$, and the product C. All calculations for the nickel-catalyzed reaction use the STO-3G minimal basis set. The upCCD state for each structure is prepared in the (8e,8o) active space and 60,365, 58,375, and 61,118 matchgate shadows are measured and used to construct the training dataset, respectively~\cite{zhao26_033061}. 

\begin{figure}[b]
    \centering
    \includegraphics[width=\columnwidth]{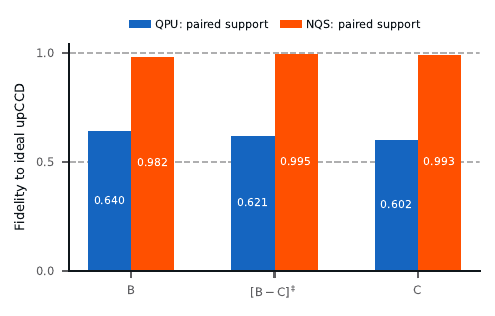}
    \caption{Fidelity to the ideal upCCD state for the direct QPU reconstruction after projection and renormalization on the 70-determinant seniority-zero space (blue) and the energy-selected shadow-trained NQS constrained to the same space (orange). The NQS trials were selected by applying the committee-fidelity filter followed by Rayleigh-energy ranking.}
    \label{fig:ni-fidelity-comparison}
\end{figure}

Figure~\ref{fig:ni-fidelity-comparison} compares the fidelity of the trained NQS with the state obtained by projecting and renormalizing the QPU shadows to the paired space. We find that the fidelity of the trained NQS ($F_{\theta}$) yields significant improvements over the projected QPU fidelities $F_{s,\mathrm{pair}}$. For all three structures, $F_{\theta}$ are close to 1 while $F_{s,\mathrm{pair}}$ could be as low as 0.60.

\begin{figure}[t]
    \centering
    \includegraphics[width=\columnwidth]{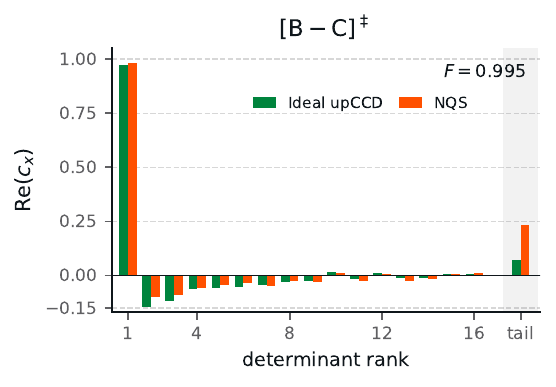}
    \caption{Coefficient-resolved comparison of the ideal upCCD state and the energy-selected NQS for $[\mathrm{B-C}]^\ddagger$.  Green and orange denote the ideal upCCD and phase-aligned NQS coefficients, respectively, with both states normalized on the 70-determinant upCCD support.  Determinants are ordered by decreasing $|c_x^{\mathrm{upCCD}}|$, with the Hartree--Fock determinant at rank one.  The selected NQS uses $\lambda =0.0545$.  The first 16 signed real coefficients are shown explicitly, and the shaded tail category gives $\sum_{i>16}|c_i|$ separately for the ideal and NQS states.  The reported fidelity is evaluated after selection using all 70 determinants.}
    \label{fig:ni-coefficients}
\end{figure}

Figure~\ref{fig:ni-coefficients} provides a determinant-resolved view of the energy-selected NQS for the transition structure $[\mathrm{B-C}]^\ddagger$.  The leading 16 determinants contain more than 99.95\% of the ideal upCCD probability, and their signed coefficients are closely reproduced by the NQS. Differences remain both within and outside the leading configurations.  The corresponding three-structure comparison is provided in Figure~\ref{fig:ni-coefficients-all-supp} of the Supplemental Information.

The improvement relative to the directly reconstructed QPU direction is consistent with error mitigation, but it is useful to be precise about what is being mitigated.  The seniority-zero constraint applied here removes amplitudes that are incompatible with the upCCD preparation, while the learned $a$ can accommodate a dataset-wide response mismatch due to finite shadow and noise-induced leakage outside of the seniority-zero space. The ridge penalty then prevents the fit from reducing the overlap error primarily through this scalar degree of freedom. Together, these restrictions favor a state direction that is substantially closer to the ideal circuit state even though that state is never used during training.  This is an ansatz-informed denoising mechanism that cannot recover components outside the assumed seniority-zero space, and it should not be interpreted as a general correction for arbitrary hardware noise.

The energies from NQS-AFQMC using the trained NQS are shown in Table~\ref{tab:nickel}. Each NQS-AFQMC calculation was propagated for 300 blocks of 20 steps, corresponding to 6,000 propagation steps and a total imaginary time of $30~E_h^{-1}$. The first 50 blocks, corresponding to $5~E_h^{-1}$, were discarded as equilibration, and the reported energies were obtained from the remaining 250-block analysis window. Besides absolute energies of the 3 structures, we also report the forward activation barrier and the signed reaction energy $\Delta E_{\mathrm{rxn}}=E_{\mathrm{C}}-E_{\mathrm{B}}$. A negative $\Delta E_{\mathrm{rxn}}$ corresponds to an exothermic reaction. We observe that NQS-AFQMC energies are in excellent agreement with those obtained using the ideal upCCD trial. More importantly, although both ph-AFQMC and  QC-AFQMC with QPU shadows predict a negative $\Delta E_{\mathrm{rxn}}$, NQS-AFQMC predicts $\Delta E_{\mathrm{rxn}}=+4.5\pm2.6$~kcal~mol$^{-1}$, consistent with the CCSD(T) reference. This agreement is consistent with mitigation of QPU errors during NQS training, although differences in propagation settings prevent us from isolating that contribution. The forward activation barrier nevertheless remains approximately $9$~kcal~mol$^{-1}$ above the CCSD(T) reference. This is due to imbalanced error cancellation between the B and the B-C structure, which is caused in part by the limitation of the upCCD trial state.

\begin{table*}[t]
\caption{Comparison of absolute energies and reaction energetics among the methods considered here. The Ideal upCCD row uses the exact active-space upCCD state as the trial wave function in multi-Slater-determinant AFQMC. NQS-AFQMC reports results obtained with the shadow-trained NQS and the workflow described in Section~\ref{sec:nqs-afqmc}. The QC-AFQMC row reports results obtained with the original QC-AFQMC implementation of Ref.~\cite{zhao26_033061}. The CCSD(T) reference energies are taken from Ref.~\cite{zhao26_033061} and were computed using the frozen-core approximation with 29 frozen orbitals.}
\label{tab:nickel}
\centering
\footnotesize
\setlength{\tabcolsep}{2.5pt}
\begin{ruledtabular}
\begin{tabular}{lccccc}
Method & \multicolumn{3}{c}{Total energy (Ha)} & \multicolumn{2}{c}{Reaction energetics (kcal mol$^{-1}$)} \\
 & B & $[\mathrm{B-C}]^\ddagger$ & C & $\Delta E^\ddagger$ & $\Delta E_{\mathrm{rxn}}$ \\
\hline
Ideal upCCD
& $-2739.081\pm0.003$ & $-2738.982\pm0.003$ & $-2739.073\pm0.003$ & $62.2\pm2.4$ & $+4.9\pm3.0$ \\
NQS-AFQMC
& $-2739.083\pm0.002$ & $-2738.984\pm0.003$ & $-2739.076\pm0.003$ & $62.6\pm2.1$ & $+4.5\pm2.6$ \\
QC-AFQMC~\cite{zhao26_033061}
& $-2739.040\pm0.004$ & $-2738.972\pm0.003$ & $-2739.059\pm0.003$ & $43\pm3$ & $-12\pm3$ \\
ph-AFQMC (RHF)~\cite{zhao26_033061}
& $-2739.074\pm0.003$ & $-2738.975\pm0.005$ & $-2739.096\pm0.005$ & $62\pm4$ & $-14\pm4$ \\
CCSD(T)~\cite{zhao26_033061}
& $-2739.081251$ & $-2738.996283$ & $-2739.068637$ & $53.3$ & $+7.9$ \\
\end{tabular}
\end{ruledtabular}
\end{table*}

\begin{table}[t]
\caption{Classical resource requirements and reaction-energy accuracy for the nickel application. Both workflows use the same QPU measurements, and the classical resource estimates are normalized to 1,500 AFQMC propagation steps for one structure. QPU cost is excluded because the measurement data are shared. The estimated classical costs use the AWS rates described in the text. The energetic deviations are defined as the result from each method minus the corresponding ideal-upCCD AFQMC result, with statistical uncertainties propagated in quadrature.}
\label{tab:nickel-cost-accuracy}
\centering
\footnotesize
\setlength{\tabcolsep}{3pt}
\begin{ruledtabular}
\begin{tabular}{lcc}
Quantity & QC-AFQMC~\cite{zhao26_033061} & NQS-AFQMC \\
\hline
QPU workload & 58,375--61,118 shadows & Same shadows \\
H200 GPU time & 13,200 h & 4.0 h \\
CPU time & Negligible & 6,680 core h \\
Estimated classical cost & \$90,600 & \$596 \\
$\delta\Delta E^\ddagger$ (kcal mol$^{-1}$) & $-19.2\pm3.8$ & $+0.4\pm3.2$ \\
$\delta\Delta E_{\mathrm{rxn}}$ (kcal mol$^{-1}$) & $-16.9\pm4.2$ & $-0.4\pm4.0$ \\
\end{tabular}
\end{ruledtabular}
\end{table}

In addition to noise resilience, the NQS also significantly reduces the classical post-processing cost of QC-AFQMC. In the direct QC-AFQMC workflow of Ref.~\cite{zhao26_033061}, every force-bias and local-energy evaluation requires an explicit sum over the $\sim 60,000$ shadows collected for each structure, for every walker. The reported postprocessing cost for a single 150-block calculation was approximately 13,200 H200 GPU hours per structure (or 1.72 days on 320 concurrent H200s). In the present workflow, this dependence on the number of shadows is incurred only once when the training overlaps are constructed. Once these data were available, training one NQS for 250 epochs required $\sim 7.5$ minutes on a single H200 GPU (or $\sim9.5$ minutes for the same calculation on an A100 GPU). The complete calibration and regularization scan contained 30 NQS optimizations for B, 30 for $[\mathrm{B-C}]^\ddagger$, and 36 for C, corresponding to an average of 32 optimizations and approximately 4.0 H200 GPU hours per structure. The shadow data is then never used during the production runs of AFQMC, 
and the cost of the production propagation therefore no longer increases when additional shadow measurements are collected. All AFQMC simulations performed for this work strictly utilize CPU resources, and the associated timings depend on the implementation. With the conservative settings chosen here, the NQS-AFQMC calculations each take approximately 19 hours of wall-clock time on an allocation of 1,280 Intel Ice Lake cores. 
The present NQS-AFQMC calculations use 6,000 propagation steps (300 blocks of 20 steps), four times the 1,500 steps used in the referenced QC-AFQMC calculations (150 blocks of 10 steps). For an equal-step comparison, we divide our timings in this work by four. Our time step is half that of the previous calculation, so an equal-projection-time comparison requires twice as many steps. Both normalizations give an estimated cost reduction of approximately 99\%. Taking all of these into account, comparison of runtime between NQS-AFQMC and QC-AFQMC is shown in Table~\ref{tab:nickel-cost-accuracy}.

As one can see, the reduction in runtime directly translates to substantial cost reduction. 
We take the p5en.48xlarge AWS instance for GPU pricing ( \$54.92 per hour for eight H200 GPUs), and the c6i.32xlarge instances for CPU pricing (\$5.44 per hour for 64 physical Intel Ice Lake cores). 
After normalizing both calculations to the same 1,500 propagation steps, the estimated cost of NQS-AFQMC is approximately \$517 per structure. We also need to include the construction of the training dataset, including the preliminary RHF-trial propagation and overlap computation with matchgate shadows, which is estimated to require an additional 600 physical core hours per structure, equivalent to approximately 1,200 AWS vCPU hours and \$51 at the same instance rate. The total CPU use is then approximately 6,680 physical core hours per structure. After taking the GPU usage of NQS training into account, the total cost for NQS-AFQMC is approximately \$596 per structure, corresponding to a reduction of approximately 99\% in estimated classical postprocessing cost relative to direct QC-AFQMC. Table~\ref{tab:nickel-cost-accuracy} summarizes this resource comparison together with the deviations in the reaction energetics from AFQMC using the ideal upCCD trial.

\section{Conclusions}\label{sec:Conclusions}

In this work, we introduced a workflow for distilling matchgate shadow measurements into a neural quantum state that can be used as a trial wave function for AFQMC. Rather than retaining explicit iterations over every shadow measurement throughout imaginary time propagation, measured shadows are first used to construct a supervised dataset of complex walker overlaps. A symmetry-constrained autoregressive NQS is trained on these data using a learnable overlap-calibration factor and ridge regularization. The resulting state is then used as the trial state to guide the imaginary time propagation of QC-AFQMC. This construction eliminates the dependence on shadow count in AFQMC post-processing. Such a dependence now only contributes toward dataset construction, which is much more efficient in time and cost. 

We demonstrate the accuracy of this approach with two examples. For the N$_2$ dissociation, the constrained NQS recovers the paired component of the trial state from hardware-measured shadows while suppressing contributions outside the intended seniority-zero space. The resulting AFQMC energies are in excellent agreement with the ideal-upCCD AFQMC energies near equilibrium, while exhibiting some limitations to fully mitigate hardware noise at the stretched geometry. 
For the nickel-catalyzed oxidative addition, the learned trials closely reproduce the ideal-upCCD AFQMC energetics and yield a reaction energy consistent with the CCSD(T) reference, suggesting that the remaining barrier errors primarily reflect limitations of the underlying upCCD trial state.

Beyond accuracy, shadow-trained NQS-AFQMC substantially reduces classical post-processing cost by confining the explicit dependence on shadow count to the one-time construction of the training dataset. Production AFQMC then uses the learned trial without revisiting the shadow measurements. For the nickel application, this gives an estimated cost reduction of approximately 99\% relative to our previous QC-AFQMC workflow~\cite{zhao26_033061}, while the resulting energetics agree more closely with the ideal-trial calculations. This reduction motivates extending the approach to larger systems.

Several questions remain concerning error mitigation. The combination of calibration, regularization, and a seniority-zero constraint is effective for the upCCD trials considered here, but its performance for more expressive trial states remains to be established. A single calibration factor may be insufficient when noise alters different wave function components nonuniformly. Extending the approach will therefore require training objectives that account for the structure of the measurement errors and the available physical constraints. Another possibility is to use the shadow-trained NQS to initialize subsequent variational energy optimization, following the broader strategy of combining data-driven initialization with Hamiltonian-based refinement~\cite{bennewitz22_618, wei24_023250}.

Scalability also requires further investigation. The largest active space studied here contains 16 qubits and remains readily accessible to classical simulation. Larger systems may require more training determinants and broader coverage of the walker distribution. It may also be useful to supplement overlap labels with force-bias or local-energy information to improve the quantities that directly govern AFQMC propagation. Such enrichment would increase the cost of constructing the dataset, making recent improvements in fermionic-shadow post-processing particularly relevant~\cite{nappi26_19239, zhao03103_2026}. Whether these improvements can support accurate surrogate training at a practical cost remains an important question.

The choice of surrogate architecture presents a further opportunity. The practical objective is to reproduce the trial-state quantities needed along AFQMC trajectories, so a globally accurate reconstruction of the quantum state is not necessarily required. An NQS provides a consistent wave function representation, but evaluating its overlaps with arbitrary Slater determinants requires an additional sampling procedure. Models that accept walker orbital matrices directly could potentially avoid this overhead, provided their overlap predictions and derived mixed estimators remain mutually consistent. Fermionic architectures inspired by approaches such as FermiNet~\cite{pfau20_033429} may offer useful design principles, although adapting them to determinant-overlap prediction would require further development.

Our results establish shadow measurements as a reusable source of training data for accelerating QC-AFQMC. By combining quantum-prepared trial states with learned classical representations, the workflow reduces repeated measurement-processing costs while retaining useful information for AFQMC propagation. Extending this approach to more expressive trials and larger systems will determine its broader role in practical quantum-classical electronic-structure calculations.

\begin{acknowledgments}
We thank Joonho Lee for related presentations ~\cite{lee25_pacifichem,lee26_qfarm} and insightful discussions with L.Z. during The International Chemical Congress of Pacific Basin Societies 2025 (Pacifichem 2025) in Dec 2025 that helped inspire this work. 
\end{acknowledgments}

\bibliography{Journal_Short_Name, ionq}

@string{ JCP = "J. Chem. Phys." }

@string{ JCTC = "J. Chem. Theory Comput." }

@string{ JPCL = "J. Phys. Chem. Lett." }

@string{ MA = "Microchim. Acta" }

@string{ MLST = "Mach. Learn.: Sci. Technol."}

@string{ N = "Nature" }

@string{ NC = "Nat. Chem." }

@string{ NMI = "Nat. Mach. Intell."}

@string{ NIPS = "Adv. Neural Inf. Process. Syst."}

@string{ NP = "Nat. Phys." }

@string{ NPJQ = "npj Quantum Inf." }

@string{ NJP = "New J. Phys." }

@string{ PRB = "Phys. Rev. B" }

@string{ PRL = "Phys. Rev. Lett." }

@string{ PRR = "Phys. Rev. Research"}

@string{ PNAS = "Proc. Natl. Acad. Sci. U.S.A." }

@string{ PICHPCNSA = "Proc. Int. Conf. High Perform. Comput. Netw. Storage Anal."}

@string{ QST = "Quantum Sci. Technol." }

@string{ S = "Science" }

@string{ WCMS = "WIREs Comput. Mol. Sci." }

@string{ Q = "Quantum"}

@article{valenti22_L012010,
  title = {{Correlation-Enhanced Neural Networks As Interpretable Variational Quantum States}},
  author = {A. Valenti and E. Greplova and N. H. Lindner and S. D. Huber},
  journal = PRR,
  volume = {4},
  pages = {L012010},
  year = {2022},
}

@article{Kiser24_033022,
  title = {{Classical and Quantum Cost of Measurement Strategies for Quantum-Enhanced Auxiliary Field Quantum Monte Carlo}},
  author = {M. Kiser and A. Schroeder and G-L. R. Anselmetti and C. Kumar and N. Moll and M. Streif and D. Vodola},
  journal = NJP,
  volume = {26},
  pages = {033022},
  year = {2024},
}

@article{Zhao20_174105,
    author = {P. R. C. Kent and A. Annaberdiyev and A. Benali and M. C. Bennet and E. J. L. Borda and P. Doak and et al},
    journal = JCP,
    title = {{QMCPACK: Advances in the Development, Efficiency, and Application of Auxiliary Field and Real-Space Variational and Diffusion Quantum Monte Carlo}},
    volume = {152},
    pages = {174105},
    year = {2020},
}

@article{Troyer21_033055,
    author = {V. von Burg and G. H. Low and T. H\"{a}ner and D. S. Steiger and M. Reiher and M. Roetteler and M. Troyer},
    title = {{Quantum Computing Enhanced Computational Catalysis}},
    journal = PRR, 
    volume = {3},
    pages = {033055},
    year = {2021},
}

@article{Garcia21_134115,
    author = {J. E. Rice and T. P. Gujarati and M. Motta and T. Y. Takeshita and E. Lee and J. A. Latone and J. M. Garcia},
    title = {{Quantum Computation of Dominant Products in Lithium–Sulfur Batteries}},
    journal = JCP,
    volume = {154},
    pages = {134115},
    year = {2021},
}

@article{Holzmann22_7001,
  title = {{Perspective on the Current State-of-the-Art of Quantum Computing for Drug Discovery Applications}},
  author = {N. S. Blunt and J. Camps and O. Crawford and R. Izs{\'a}k and S. Leontica and A. Mirani and A. E. Moylett and S. A. Scivier and C. S{\"u}nderhauf and P. Schopf and J. M. Taylor and N. Holzmann},
  journal = JCTC,
  volume = {18},
  pages = {7001-7023},
  year = {2022},
}

@article{Zhao22_60,
  title = {{Orbital-Optimized Pair-Correlated Electron Simulations on Trapped-Ion Quantum Computers}},
  author = {L. Zhao and J. Goings and K. Wright and J. Nguyen and J. Kim and S. Johri and K. Shin and W. Kyoung and J. I. Fuks and J-K. K. Rhee and Y. M. Rhee},
  journal = NPJQ,
  volume = {9},
  number = {60},
  year = {2023},
}

@article{Zhao24_76,
  title = {{Enhancing the Electron Pair Approximation with Measurements on Trapped-Ion Quantum Computers}},
  author = {L. Zhao and Q. Wang and J. J. Goings and K, Shin  and W. Kyoung and S, Noh and Y. M. Rhee and K. Kim},
  journal = NPJQ,
  volume = {10},
  number = {76},
  year = {2024},
}

@article{Umrigar143_164105,
  author = {C. J. Umrigar},
  title = {{Observations on Variational and Projector Monte Carlo Methods}},
  journal = JCP,
  volume = {143},
  pages = {164105},
  year = {2015},
}

@article{Preskill22_2022,
    author = {H-Y. Huang and R. Kueng and G. Torlai and V. V. Albert and J. Preskill},
    title = {{Provably Efficient Machine Learning for Quantum Many-Body Problems}},
    journal = S,
    volume = {377},
    pages = {1397},
    year = {2022},
}

@misc{braket,
    author = {{Braket contributors}},
    title = {{Amazon Braket: A Fully Managed Quantum Computing Service Provided by AWS}},
    year = {2019},
    version = {v1.38.0},
    url = {https://aws.amazon.com/braket}
}

@article{Zhao21_110504,
  author = {A. Zhao and N. C. Rubin and A. Miyake},
  title = {{Fermionic Partial Tomography via Classical Shadows}},
  journal = PRL,
  volume = {127},
  pages = {110504},
  year = {2021},
}

@article{huang20_1050,
  author = {H-Y. Huang and R. Kueng and J. Preskill},
  title = {{Predicting Many Properties of a Quantum System from Very Few Measurements}},
  journal = NP,
  volume = {16},
  pages = {1050-1057},
  year = {2020},
}

@article{zhang03_136401,
  author = {S. Zhang and H. Krakauer},
  title = {{Quantum Monte Carlo Method Using Phase-Free Random Walks with Slater Determinants}},
  journal = PRL,
  volume = {90},
  pages = {136401},
  year = {2003},
}

@article{lee22_7024,
  author = {J. Lee and H. Q. Pham and D. R. Reichman},
  title = {{Twenty Years of Auxiliary-Field Quantum Monte Carlo in Quantum Chemistry: An Overview and Assessment on Main Group Chemistry and Bond-Breaking}},
  journal = JCTC,
  volume = {18},
  number = {12},
  pages = {7024-7042},
  year = {2022},
}

@article{reiher17_7555,
  author = {M. Reiher and N. Wiebe and K. M. Svore and D. Wecker and M. Troyer},
  title = {{Elucidating Reaction Mechanisms on Quantum Computers}},
  journal = PNAS,
  volume = {114},
  number = {29},
  pages = {7555-7560},
  year = {2017},
}

@article{zhang97_7464,
  author = {S. Zhang and J. Carlson and J. E. Gubernatis},
  title = {{Constrained Path Monte Carlo Method for Fermion Ground States}},
  journal = PRB,
  volume = {55},
  number = {12},
  pages = {7464-7477},
  year = {1997},
}

@article{motta18_1364,
  author = {M. Motta and S. Zhang},
  title = {{Ab Initio Computations of Molecular Systems by the Auxiliary-Field Quantum Monte Carlo Method}},
  journal = WCMS,
  volume = {8},
  pages = {e1364},
  year = {2018},
}

@article{shee23_140901,
  author = {J. Shee and J. L. Weber and D. R. Reichman and R. A. Friesner and S. Zhang},
  title = {{On the Potentially Transformative Role of Auxiliary-Field Quantum Monte Carlo in Quantum Chemistry: A Highly Accurate Method for Transition Metals and Beyond}},
  journal =JCP,
  volume = {158},
  pages = {140901},
  year = {2023},
}

@article{troyer05_170201,
  author = {M. Troyer and U-J. Wiese},
  title = {{Computational Complexity and Fundamental Limitations to Fermionic Quantum Monte Carlo Simulations}},
  journal = PRL,
  volume = {94},
  pages = {170201},
  year = {2005},
}

@article{zhao26_033061,
  author = {L. Zhao and J. J. Goings and W. Aboumrad and A. Arrasmith and L. Calderin and S. Churchill and D. Gabay and T. Harvey-Brown and M. Hiles and \textit{et al}},
  title = {{Quantum-Classical Auxiliary Field Quantum Monte Carlo with Matchgate Shadows on Trapped Ion Quantum Computers}},
  journal = PRR,
  volume = {8},
  pages = {033061},
  year = {2026},
}

@article{nappi26_19239,
  title={{Quantum-Classical Auxiliary-Field Quantum Monte Carlo at the Edge of Practicability}}, 
  author={F. Nappi and M. Kiser and F. \v{S}imkovic IV},
  journal={arXiv preprint arXiv:2606.19239},
  year={2026},
}

@article{zhao03103_2026,
  author = {L. Zhao and J. Goings and E. Epifanovsky and M. Roetteler},
  title = {{Orbital-Rotation Shadow Tomography Reduces the Classical Cost of Quantum-Classical Auxiliary-Field Quantum Monte Carlo}},
  journal={arXiv preprint arXiv:2610.03103},
  year={2026},
}

@article{Claudino23_5511,
  author = {D. Claudino and  B. Peng and  K. Kowalski and T. S. Humble},
  title = {{Modeling Singlet Fission on a Quantum Computer,}},
  journal = JPCL,
  volume = {14},
  pages = {5511–5516},
  year = {2023},
}

@article{huggins22_416,
  author = {W. J. Huggins and B. A.  O'Gorman and N. C. Rubin and D. R. Reichman and R. Babbush and J. Lee},
  title = {{Unbiasing Fermionic Quantum Monte Carlo with a Quantum Computer}},
  journal = N,
  volume = {603},
  number = {7901},
  pages = {416--420},
  year = {2022},
}

@article{huang24_043063,
  author = {B. Huang and Y-T. Chen and B. Gupt and M. Suchara and A. Tran and S. McArdle and G. Galli},
  title = {{Evaluating a Quantum-Classical Quantum Monte Carlo Algorithm with Matchgate Shadows}},
  journal = PRR,
  volume = {6},
  pages = {043063},
  year = {2024},
}

@article{wan23_629,
  author = {K. Wan and W. J. Huggins and J. Lee and R. Babbush},
  title = {{Matchgate Shadows for Fermionic Quantum Simulation}},
  journal = CMP,
  volume = {404},
  number = {2},
  pages = {629-700},
  year = {2023},
}

@article{torlai18_447,
  title={{Neural-Network Quantum State Tomography}},
  author={G. Torlai and G. Mazzola and J. Carrasquilla and M. Troyer and R. G. Melko and G. Carleo},
  journal=NP,
  volume={14},
  pages={447-450},
  year={2018},
}

@article{lange25_1675,
  title={{Transformer Neural Networks and Quantum Simulators: A Hybrid Approach for Simulating Strongly Correlated Systems}},
  author={H. Lange and G. Bornet and G. Emperauger and C. Chen and T. Lahaye and S. Kienle and A. Browaeys and A.  Bohrdt},
  journal=Q,
  volume={9},
  pages={1675},
  year={2025},
}

@article{carleo17_602,
  title={{Solving the Quantum Many-Body Problem with Artificial Neural Networks}},
  author={G. Carleo and M. Troyer},
  journal=S,
  volume={355},
  number={6325},
  pages={602-606},
  year={2017},
}

@article{choo20_2368,
  title={{Fermionic Neural-Network States for Ab-Initio Electronic Structure}},
  author={K. Choo and A. Mezzacapo and G. Carleo},
  journal=NC,
  volume={11},
  pages={2368},
  year={2020},
}

@article{lange24_040501,
  title={{From Architectures to Applications: A Review of Neural Quantum States}},
  author={H. Lange and A. Van de Walle and A. Abedinnia and A. Bohrdt},
  journal=QST,
  volume={9},
  pages={040501},
  year={2024},
}

@article{hibat21_952,
  title={{Variational Neural Annealing}},
  author={M. Hibat-Allah and E. M. Inack and R. Wiersema and R. G. Melko and J. Carrasquilla},
  journal=NMI,
  volume={3},
  pages={952-961},
  year={2021},
}

@article{malyshev23_04166,
  title={{Autoregressive Neural Quantum States with Quantum Number Symmetries}},
  author={A. Malyshev and J. M. Arrazola and A. I. Lvovsky},
  journal={arXiv preprint arXiv:2310.04166},
  year={2023}
}

@article{sharir20_020503,
  title={{Deep Autoregressive Models for the Efficient Variational Simulation of Many-Body Quantum Systems}},
  author={O. Sharir and Y. Levine and N. Wies and G. Carleo and A. Shashua},
  journal=PRL,
  volume={124},
  pages={020503},
  year={2020},
}

@article{barrett22_351,
  title={{Autoregressive Neural-Network Wavefunctions for Ab Initio Quantum Chemistry}},
  author={T. D. Barrett and A. Malyshev and A. I. Lvovsky},
  journal=NMI,
  volume={4},
  pages={351-358},
  year={2022},
}

@inproceedings{zhao21_1,
  title={{Overcoming Barriers to Scalability in Variational Quantum Monte Carlo}},
  author={T. Zhao and S. De and B. Chen and J. Stokes and S. Veerapaneni},
  booktitle={{Proceedings of the International Conference for High Performance Computing, Networking, Storage and Analysis}},
  pages={1-13},
  year={2021}
}

@article{xiao25_9587,
  title={{NNQS-AFQMC: Neural Network Quantum States Enhanced Fermionic Quantum Monte Carlo}},
  author={Z-Y. Xiao and B. Kan and H. Ma and B. Zhao and H. Shang},
  journal=JCTC,
  volume={21},
  pages={9587-9600},
  year={2025},
}

@article{malone23_109,
  title={{IPIE: A Python-Based Auxiliary-Field Quantum Monte Carlo Program with Flexibility and Efficiency on CPUs and GPUs}},
  author={F. D. Malone and A. Mahajan and J. S. Spencer and J. Lee},
  journal=JCTC,
  volume={19},
  pages={109-121},
  year={2023},
}

@article{low24_08964,
  author = {Low, Guang Hao},
  title = {{Classical Shadows of Fermions with Particle Number Symmetry}},
  journal = { arXiv preprint arXiv:2208.08964},
  year = {2024},
}

@article{guo19_1957,
  title={{Nickel-Catalyzed Suzuki--Miyaura Cross-Couplings of Aldehydes}},
  author={L. Guo and W. Srimontree and C. Zhu and B. Maity and X. Liu and L. Cavallo and M. Rueping},
  journal=NC,
  volume={10},
  pages={1957},
  year={2019},
}

@article{paszke19_8026,
  title={{Pytorch: An Imperative Style, High-Performance Deep Learning Library}},
  author={A. Paszke and S. Gross and F. Massa and A. Lerer and J. Bradbury and G. Chanan and T. Killeen and Z. Lin and N. Gimelshein and L. Antiga and \textit{et al}},
  journal = NIPS,
  volume = {32},
  pages = {8026-8037},
  year={2019}
}

@article{zhao23_025034,
  title={{Scalable Neural Quantum States Architecture for Quantum Chemistry}},
  author={T. Zhao and J. Stokes and S. Veerapaneni},
  journal=MLST,
  volume={4},
  pages={025034},
  year={2023},
}

@article{bennewitz22_618,
  title={{Neural Error Mitigation of Near-Term Quantum Simulations}},
  author={E. R. Bennewitz and F. Hopfmueller and B. Kulchytskyy and J. Carrasquilla and P. Ronagh},
  journal=NMI,
  volume={4},
  pages={618-624},
  year={2022},
}

@inproceedings{wu23_1,
  title={{NNQS-Transformer: An Efficient and Scalable Neural Network Quantum States Approach for Ab Initio Quantum Chemistry}},
  author={Y. Wu and C. Guo and Y. Fan and P. Zhou and H. Shang},
  booktitle=PICHPCNSA,
  pages={1-13},
  year={2023}
}

@article{shang25_8464,
  title={{Solving the Many-Electron {Schr\"odinger} Equation with a Transformer-Based Framework}},
  author={H. Shang and C. Guo and Y. Wu and Z.  Li and J. Yang},
  journal=NC,
  volume={16},
  pages={8464},
  year={2025},
}

@article{knitter25_025022,
  title={{Retentive Neural Quantum States: Efficient Ans{\"a}tze for Ab Initio Quantum Chemistry}},
  author={O. Knitter and D. Zhao and J. Stokes and M. Ganahl and S. Leichenauer and S. Veerapaneni},
  journal=MLST,
  volume={6},
  pages={025022},
  year={2025},
}

@article{yang20_012039,
  title={{Deep Learning-Enhanced Variational Monte Carlo Method for Quantum Many-Body Physics}},
  author={L. Yang and Z. Leng and G. Yu and A. Patel and W-J. Hu and H. Pu},
  journal=PRR,
  volume={2},
  pages={012039},
  year={2020},
}

@article{wei24_023250,
  title={{Neural-Shadow Quantum State Tomography}},
  author={V. Wei and W. A. Coish and P. Ronagh and C. A. Muschik},
  journal=PRR,
  volume={6},
  pages={023250},
  year={2024},
}

@article{pfau20_033429,
  title = {{Ab Initio Solution of the Many-Electron Schr\"odinger Equation with Deep Neural Networks}},
  author = {D. Pfau and J. S. Spencer and A. D. G. Matthews and W. M. C. Foulkes},
  journal = PRR,
  volume = {2},
  pages = {033429},
  year = {2020},
}

@misc{lee25_pacifichem,
  title={{Ab-initio Auxiliary-Field Quantum Monte Carlo Made Better with Quantum Computers}},
  author={J. Lee},
  howpublished={{Oral presentation at the 2025 International Chemical Congress of Pacific Basin Societies (Pacifichem 2025)}},
  note={{Honolulu, Hawaii, December 18, 2025}},
  year={2025},
}

@misc{lee26_qfarm,
  title={{Q-FARM Seminar}},
  author={J. Lee},
  howpublished={{Stanford University}},
  url={https://qfarm.stanford.edu/events/q-farm-seminars/joonho-lee-harvard},
  note={{February 4, 2026}},
  year={2026},
}
\clearpage

\appendix

\section{Determinant-dependent and uniform calibration}
\label{app:calibration}

In the main text, we use a single positive scalar $a$ to account for the difference in scale between the normalized NQS and the shadow-derived overlap labels. Here, we show how this approximation follows from a more general determinant-dependent response and identify a simple noise model for which the uniform response is exact.

\subsection{Determinant-dependent response}

In a general diagonal response model, each configuration can have its own nonnegative calibration factor. Let $m$ label active-space walker determinants $\ket{\Phi_m}$ and let $\ket{x_i}$ enumerate the configurations allowed by the NQS. The predicted overlap is then
\begin{equation}
    \hat{y}_m
    =\sum_i \langle\Phi_m|x_i\rangle
      a_i\psi_\theta(x_i)+\epsilon_m,
    \qquad a_i\geq0.
    \label{eq:determinant-dependent-response}
\end{equation}
The factors $a_i$ enter inside the determinant sum because each walker overlap contains contributions from many configurations. Priors centered at $a_0^{(i)}$ could incorporate independent calibration data or known relationships among determinants. However, the overlap data primarily determine the products $a_i\psi_\theta(x_i)$. Learning every $a_i$ together with the NQS amplitudes is therefore not identifiable without additional information. Relative phase errors and mixing among determinants would require a more general response than the nonnegative diagonal model in Eq.~\eqref{eq:determinant-dependent-response}.

\subsection{Uniform-response approximation}
\label{app:calibration-uniform}

We consider the simplest limit in which the dominant scale mismatch is common to every configuration. Setting $a_i=a$ gives
\begin{equation}
    \sum_i \langle\Phi_m|x_i\rangle a\psi_\theta(x_i)
    =a\,\langle\Phi_m|\Psi_\theta\rangle.
    \label{eq:uniform-response-reduction}
\end{equation}
To obtain the training objective, we take the residual errors to be circular complex Gaussian variables and place a Gaussian prior on $a$ centered at $a_0$ and restricted to the nonnegative real axis,
\begin{equation}
    \pi(a)\propto
    \exp\!\left[-\frac{(a-a_0)^2}{2\sigma_a^2}\right]
    \Theta(a),
    \label{eq:uniform-response-prior}
\end{equation}
where the Heaviside step function $\Theta(a)=1$ for $a\geq0$ and is zero otherwise. Up to an overall scale and terms independent of $a$ and $\theta$, the corresponding negative log posterior is
\begin{equation}
\begin{aligned}
    \mathcal{L}(\theta,a)
    &= \frac{1}{N_D}\sum_{m=1}^{N_D}
      \left|a\,\langle\Phi_m|\Psi_\theta\rangle-\hat{y}_m\right|^2\\
    &\quad+\lambda(a-a_0)^2,
    \qquad
    \lambda=\frac{\sigma_\epsilon^2}{2N_D\sigma_a^2}.
\end{aligned}
    \label{eq:uniform-map-objective}
\end{equation}
Equation~\eqref{eq:calibrated-shadow-loss} follows from the zero-centered choice $a_0=0$. A nonzero center can instead encode independent information about the expected scale. For example, if the labels contain only the projection of $\ket{\Psi_T}$ onto the NQS support $\mathcal S$, the corresponding ideal scale is $a_0=\sqrt{\langle\Psi_T|P_{\mathcal S}|\Psi_T\rangle}$. In the presence of finite-shadow error or contributions outside $\mathcal S$, the learned $a$ should instead be regarded as a fitted scale for the complete training set.

\subsection{Noise models with a uniform response}
\label{app:calibration-noise}

The uniform model has a direct physical interpretation under global depolarizing noise. Consider a channel that acts on the complete reference state in Eq.~\eqref{eq:bell-reference-state} on $n$ occupation qubits,
\begin{equation}
    \mathcal E_\mathrm{global}(\rho)=(1-p)\rho+p\frac{I}{2^n}.
    \label{eq:calibration-depolarizing-channel}
\end{equation}
Because the identity has no coherence between the vacuum and a nonzero-particle sector, the resulting vacuum-to-$\eta$-particle coherence is
\begin{equation}
    \ket{\Psi_{\text{eff}}}\equiv2P_\eta\mathcal E_\mathrm{global}(\rho_\tau)\ket{\mathrm{vac}}
    =(1-p)\ket{\Psi_T},
    \label{eq:calibration-global-depolarization}
\end{equation}
where $\rho_\tau=\ket{\tau}\!\bra{\tau} = \frac{1}{2}(\ket{\text{vac}} + \ket{\Psi_T})(\bra{\text{vac}} + \bra{\Psi_T})$ and $P_\eta$ projects onto the particle number sector of $\ket{\Psi_T}$. The mean overlap with a relevant determinant is therefore reduced by the same factor, and the uniform calibration is exact with $a=1-p$. This parallels the observation in infidelity-based shadow training that gate-independent noise only rescales the loss gradient~\cite{wei24_023250}. A regression on overlap labels is not scale-invariant, so the scale must be calibrated explicitly.

The same uniform response is obtained for identical single-qubit channels applied independently to all $n$ qubits that represent spin-orbital occupancy. We use the following conventions for dephasing, amplitude damping, and local depolarizing noise, respectively:
\begin{equation}
    \begin{aligned}
        \mathcal E_{\mathrm{deph}}(\rho)
        &=\frac{1+\kappa}{2}\rho+\frac{1-\kappa}{2}Z\rho Z,\\
        \mathcal E_{\mathrm{damp}}(\rho)
        &=K_0\rho K_0^\dagger+K_1\rho K_1^\dagger,\\
        \mathcal E_{\mathrm{local}}(\rho)
        &=(1-p_\ell)\rho+p_\ell\operatorname{Tr}(\rho)\frac{I}{2}.
    \end{aligned}
    \label{eq:calibration-single-qubit-channels}
\end{equation}
Here, $Z = \ket{0}\!\bra{0}-\ket{1}\!\bra{1}$, $K_0=\ket{0}\!\bra{0}+\sqrt{1-\gamma}\ket{1}\!\bra{1}$, and $K_1=\sqrt{\gamma}\ket{0}\!\bra{1}$, where $\kappa$ is the dephasing factor, $\gamma$ is the amplitude damping probability, and $p_\ell$ is the local depolarizing probability. For a determinant $\ket{x}$ with $n$ orbitals and $\eta$ electrons, the reference coherence $\ket{x}\!\bra{\mathrm{vac}}$ contains $\eta$ local factors of $\ket{1}\!\bra{0}$ and $n-\eta$ factors of $\ket{0}\!\bra{0}$. The three channels multiply each occupied orbital coherence, $\ket{1}\!\bra{0}$, by $\kappa$, $\sqrt{1-\gamma}$, and $1-p_\ell$, respectively. Dephasing and amplitude damping leave the unoccupied orbital factor unchanged, whereas local depolarization contributes an additional factor of $1-p_\ell/2$ for each unoccupied orbital. Therefore, if we instead apply these channels in Eq.~\ref{eq:calibration-global-depolarization} and write $\ket{\Psi_{\text{eff}}}=a\ket{\Psi_T}$, we find that
\begin{equation}
    \begin{aligned}
        a_{\mathrm{deph}}&=\kappa^\eta,\\
        a_{\mathrm{damp}}&=(1-\gamma)^{\eta/2},\\
        a_{\mathrm{local}}&=(1-p_\ell)^\eta(1-p_\ell/2)^{n-\eta}.
    \end{aligned}
    \label{eq:calibration-local-noise-models}
\end{equation}
These expressions give the exact uniform calibration factor for the extracted reference coherence when the trial state has fixed particle number and the channel parameters are the same on every qubit.

\clearpage
\setcounter{figure}{0}
\renewcommand{\thefigure}{S\arabic{figure}}
\renewcommand{\theHfigure}{S\arabic{figure}}

\section{Supplemental information: Noise-model shadow reconstruction}
\label{app:n2-noise-shadow}

\begin{figure*}[t]
    \centering
    \includegraphics[width=\textwidth]{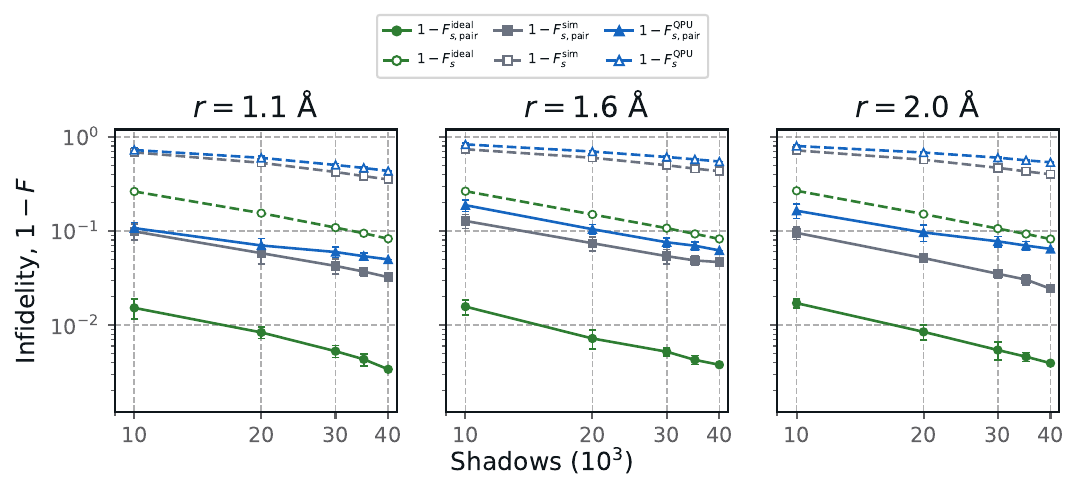}
    \caption{Supplemental comparison of the matchgate shadow reconstruction infidelity, $1-F$, for N$_2$ at $r=1.1$, $1.6$, and $2.0~\text{\AA}$. Solid lines and filled markers denote the paired support, while dashed lines and open markers denote the complete sector. Points below 40,000 shadows are means over ten matched random subsets sampled without replacement, and error bars give one standard deviation.}
    \label{fig:n2-shadow-fidelity-all}
\end{figure*}

For completeness, Fig.~\ref{fig:n2-shadow-fidelity-all} supplements the ideal-simulator and QPU comparison in Fig.~\ref{fig:n2-shadow-fidelity} with matched simulations using the Forte Enterprise noise model. The same matchgate circuits and randomly selected subsets are used for each available source within a given geometry. At 40,000 shadows, the noise-model fidelities on the paired support range from 0.953 to 0.976, while the corresponding full-sector fidelities range from 0.567 to 0.647. The noise-model results therefore reproduce the qualitative separation between accurate reconstruction within the paired manifold and the lower fidelity obtained after normalization over the complete particle-number and spin sector.

\section{Supplemental information: Nickel determinant coefficients}
\label{app:ni-determinant-coefficients}

\begin{figure*}[t]
    \centering
    \includegraphics[width=\textwidth]{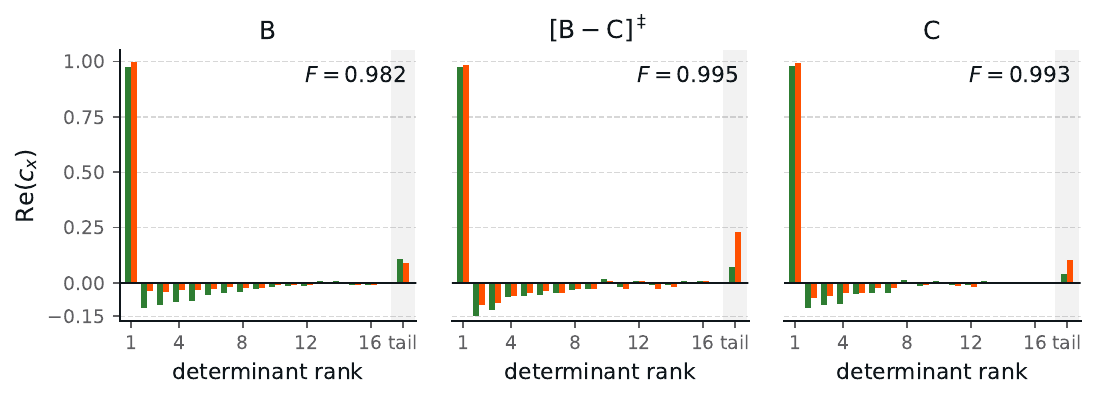}
    \caption{Coefficient-resolved comparison of the ideal upCCD states and the corresponding NQS trials used for the production AFQMC results in Table~\ref{tab:nickel} for B, $[\mathrm{B-C}]^\ddagger$, and C. Green and orange denote the ideal upCCD and phase-aligned NQS coefficients, respectively, with all states normalized on the 70-determinant upCCD support. Within each structure, determinants are ordered by decreasing $|c_x^{\mathrm{upCCD}}|$, with the Hartree--Fock determinant at rank one. The first 16 signed real coefficients are shown explicitly, and the shaded tail category gives $\sum_{i>16}|c_i|$ separately for the ideal and NQS states.}
    \label{fig:ni-coefficients-all-supp}
\end{figure*}

Figure~\ref{fig:ni-coefficients-all-supp} extends the determinant-resolved comparison in Fig.~\ref{fig:ni-coefficients} to all three structures along the oxidative-addition pathway.

\end{document}